\documentclass[conference]{IEEEtran}

\pdfoutput=1
\usepackage{amsmath,amsfonts,amssymb}
\usepackage{algorithm,algorithmic}
\usepackage{listings,enumerate, graphicx, epsfig, cite, subfigure}
\usepackage{multirow,color}

\begin{document}
\IEEEoverridecommandlockouts
\title{Channel Protection: Random Coding Meets Sparse Channels}
\date{\today \\ at \currenttime}

\author{
\IEEEauthorblockN{M. Salman Asif, William Mantzel and Justin Romberg}
\IEEEauthorblockA{School of Electrical and Computer Engineering \\
Georgia Institute of Technology,
Atlanta, Georgia, 30332, USA
\\ Email: \{sasif, willem, jrom\}@ece.gatech.edu} }


\maketitle

\begin{abstract}
Multipath interference is an ubiquitous phenomenon in modern communication
systems. The conventional way to compensate for this effect is to equalize the
channel by estimating its impulse response by transmitting a set of
training symbols. The primary drawback to this type of approach is
that it can be unreliable if the channel is changing rapidly.
In this paper, we show that randomly encoding the signal can protect it against channel uncertainty when the channel is sparse. Before transmission, the signal is mapped into a slightly longer codeword using a random matrix.
From the received signal, we are able to simultaneously estimate the
channel and recover the transmitted signal.
We discuss two schemes for the recovery. Both of them exploit the
sparsity of the underlying channel. We show that if the
channel impulse response is sufficiently sparse, the transmitted signal can be
recovered reliably.
\end{abstract}
\section{Introduction}
In many communication systems, transmitted signals suffer from multipath interference \cite{Goldsmith-2005-Wireless}. These effects can be mitigated on the receiver end by estimating the channel response, either by using training (or pilot) symbols \cite{Cavers-1991-PilotAnalysis} or by blind equalization \cite{Liu-1996-Recent}. In a number of applications, the channel exhibits a sparse impulse response, i.e. a small nonzero support. In such cases, the sparse nature of the channel can be exploited to get better estimates \cite{CotterRao_2002_SparseChannelEst}. In recent years several training based methods have been presented in this regard \cite{CotterRao_2002_SparseChannelEst, Cetin-2005-Semi, carbonelli2007sparse,Li2007estimation,Bajwa_2008_CsChannelEst}. However, if the channel response is unknown, partially known, and/or changing rapidly, some other mechanism is needed to protect the transmitted signals from the channel interference.

In this paper we demonstrate how a \emph{random coding} scheme can protect an arbitrary signal from the effects of sparse channels. We show that if the original signal is encoded using a random matrix before transmission, then we can estimate the channel and recover the signal exactly (when there is no noise present) on the receiver side. Our scheme does not use any prior knowledge of the channel, other than that the impulse response of the channel is sparse.

The problem can be formulated as follows. We want to transmit a signal $x\in \mathbb{R}^n$ to a remote receiver. In order to protect the signal from the multipath interference, we instead transmit a codeword $Ax$, where $A$ is an $m\times n$ random coding matrix with $m > n$. The received signal can be expressed as
\begin{equation}\label{eq:y=Ax*h}
y = Ax\otimes h,
\end{equation}
where $h \in \mathbb{R}^m$ is a $k$-sparse vector (i.e., contains only $k$ nonzero entries) for the impulse response of the channel and  $\otimes$ denotes the circular convolution. Our goal is to recover the original signal $x$ from the received signal $y$.

Few comments about the problem formulation are in order. The coding matrix $A$ can be generated by drawing its entries independently from a subgaussian distribution, e.g., Gaussian or Bernoulli \cite{Baraniuk-2008-JLRIP}. However, in this paper, we assume that the coding matrix $A$ is Gaussian, whose entries are independently drawn from the normal distribution $N(0,1/m)$. Although we use circular convolution in \eqref{eq:y=Ax*h}, the setup and subsequent discussion can be easily extended to the linear convolution. With our assumption that $h$ is sparse, for the recovery we can use methods from sparse signal approximation literature \cite{Chen_99_BasisPursuit}, in particular, compressive sensing \cite{CRT06_RobustUP, Donoho_2006_CS}.

In general, compressive sensing deals with the problems involving the recovery of sparse signals from small number of linear measurements. Consider the following setup. Assume that we want to recover a sparse signal $\alpha \in \mathbb{R}^n$ from its linear measurements $b=\Phi \alpha$, where $\Phi$ is an $m\times n$ matrix with $m\ll n$. Ideally, we would like to solve the following optimization problem
\begin{equation}\label{eq:L0}
\text{minimize} \; \|\tilde \alpha\|_0 \;\; \text{subject to} \;\; b = \Phi \tilde \alpha,
\end{equation}
where $\|\tilde \alpha\|_0$ denotes the $\ell_0$ norm (i.e., cardinality) of the optimization variable $\tilde \alpha$. Unfortunately, the problem in \eqref{eq:L0} is combinatorial in nature, and known to be NP-hard \cite{Natarajan_1995_SparseApprox}. However, a relaxation of $\ell_0$ norm with $\ell_1$ norm makes the optimization problem convex, and can be solved efficiently \cite{Boyd_book_ConvexOptimization}.
The resultant $\ell_1$ optimization problem (also known as basis pursuit) can be written as
\begin{equation}\label{eq:BP}
\text{minimize} \; \|\tilde \alpha\|_1 \;\; \text{subject to } \;\; b = \Phi \tilde \alpha.
\end{equation}
If the original signal $\alpha$ is $k$-sparse and $\Phi$ obeys some \emph{incoherence} conditions (e.g., $\Phi$ is Gaussian or Bernoulli), then the solution of \eqref{eq:BP} is exact (with high probability) if $m = O(k \log (n/k))$ \cite{CandesTao_2006_NearOptimal}.

A related problem in compressive sensing, where random coding is used, is the $\ell_1$ decoding -- an error correction scheme \cite{CandesTao_2005_DecodingLP}. In this framework we use random coding to protect an arbitrary signal from the ``sparse additive'' errors. The problem setup is as follows. Assume that we want to reliably transmit a signal $x \in \mathbb{R}^n$ to the receiver. In order to do so, we transmit a codeword $Ax$, where $A$ is an $m\times n$ random coding matrix with $m>n$. The received signal is
\begin{equation}\label{eq:y=Ax+e}
y = Ax+e,
\end{equation}
where $e\in \mathbb{R}^m$ is the sparse error introduced by the channel. If $e$ contains $k$ nonzero terms and $m-n = O(k \log (m/k))$, then solving the following optimization program will recover $x$ exactly (with very high probability) \cite{CandesTao_2005_DecodingLP, rudelson05ge}
\begin{equation}\label{eq:L1-Decoding}
\text{minimize} \; \|A\tilde x - y\|_1.
\end{equation}

In this paper we use random coding to protect an arbitrary signal from the effects of a ``sparse convolutive'' channel. We present two schemes for the signal recovery. The first scheme formulates the recovery problem as a block sparse signal recovery problem. The second scheme is an improvement over block sparse method and it uses an alternating minimization algorithm to recover $x$ and $h$. We also present some simulation results to demonstrate the performance of random coding. 

\section{Block $\ell_1$}\label{sec:Block-L1}
In this section we discuss the block-sparse formulation to estimate $x$ and $h$ in \eqref{eq:y=Ax*h}. The system in \eqref{eq:y=Ax*h} can be written as
\begin{equation}\label{eq:y=sum_hAx}
y = \sum_{i=1}^{m} h(i) S^{i-1} A x,
\end{equation}
where $S^i$ denotes an operator for $i$ circular downward shifts of the columns, $h(i)$ is the $i$th element of $h$. By noting that $h(i) x$ is simply the $i$th column of the rank-$1$ matrix $U=x h^T \in \mathbb{R}^{n \times m}$, we may rewrite this as:
\begin{equation}\label{eq:y=calAx}
y = \mathcal{A} \vec{U},
\end{equation}
where the $\vec{U}$ is produced from the vectorization operation $\vec{U} = [U(1,1),~U(2,1),~...,~U(n,m)]^T$, and $\mathcal{A}$ is an $m \times mn$ matrix with $n$ circular shifted copies of $A$ stacked together: $\mathcal{A} = [A~~ S^1 A~~ S^2 A ~~ \cdots ~~ S^{m-1}A]$. This vector $\vec{U}$ is called {\em block-sparse} because the nonzero elements occur in particular contiguous regions (or blocks) of $n$ elements inside this vector corresponding to the nonzero columns $U_i = h(i) x$ ($\forall ~i:  h(i) \ne 0$).

Knowing that $\vec{U}$ is $kn$-sparse, we could solve for $\vec{U}$ using the standard $\ell_1$ minimization \eqref{eq:BP}. However, the block-sparse form lends itself to much more sophisticated approaches that are possible when the sparsity has a structured model \cite{Baraniuk_2008_ModelCS}. In block-sparse signals, the nonzero entries are clustered together \cite{Stojnic_2008_OnBlockSparse}. This way, if $h$ is $k$-sparse, we may consider $\vec{U}$ as a $k$-block-sparse signal and solve the following block $\ell_1$ optimization program to estimate $\vec{U}$ from $y$
\begin{equation}\label{eq:Block-L1}
\text{minimize} \; \sum_{i=1}^m \| U_i\|_2 \;\; \text{subject to}\;\; y = \mathcal{A} \vec{U}.
\end{equation}
This is a convex program and can be solved as a second order cone program or semi definite program \cite{Boyd_book_ConvexOptimization}. The performance guarantees for the block-sparse recovery have been presented in \cite{Stojnic_2008_OnBlockSparse,Blumensath_2008_SamplingUnion,EldarMishali_2008_Robust}.

It follows from recent results in compressive sensing theory that with a {\em block  Gaussian circulant} matrix $\mathcal{A}$ as in \eqref{eq:y=calAx}, the solution of \eqref{eq:Block-L1} is exact with high probability if $m = O(kn \log^5(mn))$ \cite{Neelamani-2009-ChannelSeparation, Romberg-2009-ChannelSeparation}. This means that in order to recover $x$ exactly at the receiver side, we need $n$ elements in codeword for every single element in the channel impulse response. This multiplicative relation between the lengths of $x$ and codeword $Ax$ makes the block $\ell_1$ scheme unattractive. However, this bound on the required length of the codeword is not optimal. Since there are only $n+k$ unknown elements in \eqref{eq:y=Ax*h}, we would like to recover $x$ and $h$ with $m = O((n+k)\log(m))$.

Although the block $\ell_1$ approach improves upon na\"\i ve $\ell_1$ minimization, there are still some fundamental limitations of the block-sparse recovery method that hinder its performance. In particular, this formulation lacks the ``rank one'' constraint on the desired block-sparse solution $U$. Ideally we would like to solve the following optimization problem
\begin{align*}
\text{minimize} \; &\sum_{i=1}^m \|U_i\|_2 \\ \text{subject to}\;\; &y = \mathcal{A} \vec{U}\\
&\text{rank}(U) = 1.
\end{align*}
Indeed, this rank constraint would critically reduce the number of degrees of freedom from $O(nk)$ to $O(n+k)$, enabling solutions with substantially fewer measurements. However, this constraint is a non-convex problem and solving such a program is known to be NP-hard \cite{Recht-2007-Guaranteed_RankMin}. This rank constraint can be relaxed (e.g. with a nuclear norm constraint), and we discuss such approaches in another paper \cite{MAR-2009-CP_Allerton}.

\section{Alternating minimization}\label{sec:AM}
In this section we discuss an alternating minimization algorithm to recover  $x$ and $h$ from the measurements $y$ in \eqref{eq:y=Ax*h}. Before getting into the discussion about alternating minimization, consider the following two simpler problems. Suppose we already know the channel response $h$ and we want to estimate $x$ from $y$. In this case we can write the system in \eqref{eq:y=Ax*h} as
\begin{equation}\label{eq:y=Hx}
y = Hx,
\end{equation}
where $H = h\otimes A = \sum_{i=1}^m h(i)S^{i-1}A$ is an $m\times n$ matrix whose columns are circular convolution of $h$ with the columns of $A$. Assuming that $H$ has full column rank, $x$ can be computed exactly by solving the following least squares problem
\begin{equation}\label{eq:LS}
\text{minimize} \; \|H\tilde x - y\|_2^2.
\end{equation}
We can write $H= F \Sigma_{\hat h} F^* A$, where $F^*$ denotes a DFT matrix and $\Sigma_{\hat h}$ is a diagonal matrix consisting of $\hat h = F^* h$. In order to recover $x$ exactly (using the perfect knowledge of $h$), the rank of matrix $H$ should be at least $n$. This implies that ${\hat h}$ must have at least $n$ non-zero entries. The discrete uncertainty principle \cite{Donoho_1989_UP} tells us that for any $m$-dimensional signal $h$ and its Fourier transform $\hat h$, the following relation holds for the time and frequency support sizes
\[\|h\|_0 + \|\hat h\|_0 \ge 2\sqrt{m}. \]
The quantitative robust uncertainty principle \cite{Candes_2006_QuantitativeRobustUP} improves on this bound and suggests that, for \emph{almost} every $h$, the following relation holds for the time and frequency support sizes
\[\|h\|_0 + \|\hat h\|_0 \gtrsim \frac{m}{\log m}.\]
Since we definitely need $\|\hat h\|_0 \ge n$, a minimal bound on $m$ can be derived as:
\[\|\hat h\|_0 \gtrsim \frac{m}{\log m} - k \ge n \quad \Rightarrow \quad m \gtrsim (n+k)\log m.\]
Therefore, even when we know the channel response $h$ exactly, to reliably recover $x$ in \eqref{eq:y=Hx} we need $m \gtrsim (n+k)\log m$.

On the other hand, suppose we know $x$ and we want to estimate $h$ from $y$. The system in \eqref{eq:y=Ax*h} can be written as
\begin{equation}\label{eq:y=Ch}
y = Ch,
\end{equation}
where $C$ is the circulant matrix created from the vector $Ax$. The sparse vector $h$ can be estimated using any standard sparse recovery method. Since $A$ is Gaussian matrix and $x$ is a deterministic signal, $Ax$ is a Gaussian vector and $C$ is a Gaussian circulant matrix. Therefore, (with the complete knowledge of $x$) we can recover $h$ exactly by solving the basis pursuit problem if $m \gtrsim k \log^5(m)$ \cite{Romberg-2009-UUP_Circulant}.

Now consider the original problem, where we do not know $x$ and $h$, and the only prior information we have is that $h$ is sparse. In this regard, we wish to solve the following joint optimization problem for $x$ and $h$
\begin{equation}\label{eq:AM-x,h}
\underset{\tilde x, \tilde h}{\text{minimize}} \; \frac{1}{2}\|\tilde h \otimes A \tilde x - y\|_2^2+\tau \|\tilde h\|_1,
\end{equation}
where $\tau>0$ is a regularization parameter. There are two costs involved in \eqref{eq:AM-x,h}. The quadratic term in \eqref{eq:AM-x,h}, known as the residual, tries to keep estimates close to the measurements. The $\ell_1$ norm works towards making the estimate of channel sparse. Note that for a fixed value of $h$, \eqref{eq:AM-x,h} reduces to the least squares problem \eqref{eq:LS}. On the other hand, for a fixed value of $x$ \eqref{eq:AM-x,h} becomes convex in $h$, and takes the form of a well studied sparse recovery problem -- basis pursuit denoising (BPDN) \cite{Chen_99_BasisPursuit}. However, the optimization problem \eqref{eq:AM-x,h} is not convex in both $x$ and $h$ simultaneously, which makes it a challenging task to solve this problem. We use the alternating minimization scheme to solve \eqref{eq:AM-x,h} for $x$ and $h$.

The alternating minimization (also known as projection method) is a commonly used method to solve an optimization problem over two variables \cite{Csiszar-1984-InfoGeomAM}. In the framework of alternating minimization, instead of solving the optimization problem over multiple variables simultaneously, we solve a sequence of optimization problems over one variable at a time. This scheme of ``alternating the minimization variables'' has been used in a variety of applications. Examples include blur identification and image restoration, blind source separation, adaptive filter design and  matrix factorization.

The alternating minimization (AM) algorithm for \eqref{eq:AM-x,h} is an iterative scheme. Every AM iteration can be divided into two main steps, where we alternately minimize \eqref{eq:AM-x,h} over $h$ and $x$. We use subscript $j$ to denote the variables at $j$th iteration of the algorithm. Assume that $x_{j-1}$ and $h_{j-1}$ are the solutions at the beginning of $j$th iteration of AM. The two main steps of the AM iteration can be described as follows.
\begin{enumerate}
\item With the signal $x$ fixed as $x_{j-1}$, update the channel estimate $h_j$ by solving \eqref{eq:AM-x,h} over $h$ as
    \begin{equation}\label{eq:AM-step1}
    \underset{}{\text{minimize}} \; \frac{1}{2} \|C_j \tilde h - y\|_2^2 + \tau_j \|\tilde h\|_1,
    \end{equation}
    where $\tau_j$ is the regularization parameter and $C_j$ is the circulant matrix created from the vector $Ax_{j-1}$.
\item With channel $h$ fixed as $h_j$, update the signal estimate $x_j$ by solving \eqref{eq:AM-x,h} over $x$ as
    \begin{equation}\label{eq:AM-step2}
    \underset{}{\text{minimize}} \; \|H_j\tilde x - y\|^2_2,
    \end{equation}
    where $H_j = h_j \otimes A = \sum_{i=1}^m h_j(i) S^{i-1}A$.
\end{enumerate}
Repeat the procedure of updating $h_j$ and $x_j$ until some convergence criterion is satisfied.

\algsetup{indent = 1em}
\begin{algorithm}[t]
\caption{Alternating minimization for channel protection}\label{alg:AM}
\begin{algorithmic}
\STATE Start with the initial solutions $x_0$ and $h_0$ for \eqref{eq:AM-x,h} at $j=0$.
\REPEAT
\STATE $ j \leftarrow j+1$
\STATE Select the value of $\tau_j$ (or the cardinality $k_j$ for the channel estimate).
 \STATE Fix $x = x_{j-1}$ in \eqref{eq:AM-x,h} and solve it as \eqref{eq:AM-step1} to compute $h_j$.
\STATE Fix $h = h_j$ in \eqref{eq:AM-x,h} and solve it as \eqref{eq:AM-step2} to compute $x_j$.
\STATE Normalize $x_j$ as $x_j = \dfrac{x_j}{\|x_j\|_2}$.
\UNTIL{\texttt{stopping criterion is satisfied}}
\end{algorithmic}
\end{algorithm}

A sketch for the AM algorithm is given in Algorithm~\ref{alg:AM}. Here we discuss some important features of the AM algorithm in detail.\\
\textbf{Nonconvexity: } The optimization problem \eqref{eq:AM-x,h} is nonconvex, and it allows several possible solutions (i.e., the local minima). The final solution where Algorithm~\ref{alg:AM} converges depends on several factors. The most important factors in this regard are the choices for the initial values of $x_0$ and $h_0$ and the value of regularization parameter $\tau_j$ at every iteration. There can be many ways to initialize the Algorithm~\ref{alg:AM} and then control the regularization parameter $\tau_j$, so that the Algorithm~\ref{alg:AM} converges to the global minimum for \eqref{eq:AM-x,h} (which is often at the correct solution). In the following discussion, we restrict ourselves to one such setup, where we start with an initial channel estimate which has only one nonzero entry and slowly add more entries in the channel estimate as we move towards the final solution. \\
\textbf{Initialization: } We start Algorithm~\ref{alg:AM} by identifying the strongest path (component) in the channel response. The initial channel estimate $h_0$ is selected as a vector which is one at the location corresponding to the strongest path and zero everywhere else. This way the initial estimate of the channel becomes just a delay function. 
The initial value for $x_0$ can then be computed as the least squares estimate, using $h_0$ in \eqref{eq:AM-step2}. 
The initial value of channel estimate $h_0$ can either be known as a prior knowledge (e.g., the line of sight path is dominant) or it can be estimated from the received signal $y$ as follows.
We can identify the strongest path in the channel, and consequently $x_0$ and $h_0$, by solving the following least squares problem for every possible delay $i$ as
\begin{equation}\label{eq:Initial_x0}
\underset{\tilde x}{\text{minimize}} \; \|S^{i-1} A \tilde x - y\|_2^2 \quad \text{for }  i\in \{1,\ldots, m\}.
\end{equation}
The delay which gives the smallest residual in \eqref{eq:Initial_x0} corresponds to the strongest path in the channel. This ultimately gives us the values for the initial estimates $x_0$ and $h_0$. This scheme of finding the strongest path can be considered as a generalized matched filter.
\\
\textbf{Selection of $\tau_j$ or cardinality $k_j$: } The next important task is the selection of the regularization parameter $\tau_j$ in \eqref{eq:AM-step1}. At any  $j$th iteration of Algorithm~\ref{alg:AM}, we wish to solve \eqref{eq:AM-step1} such that the solution $h_j$ has cardinality $k_j$ (i.e., number of nonzero entries). We slowly increase $k_j$ as we proceed towards the final solution, e.g., $k_j = \left\lceil \dfrac{j}{r}\right\rceil$ for some constant $r\ge 1$. The value of $\tau_j$ in \eqref{eq:AM-step1}, to a large extent, controls the cardinality of the solution $h_j$. For example, $h_j$ is zero if $\tau_j > \|C_j^T y\|_\infty$, and nonzero entries start to appear in $h_j$ as we reduce $\tau_j$. However, it is difficult to predict the value of $\tau_j$ which gives a $k_j$-sparse solution of \eqref{eq:AM-step1}. Fortunately, we have a homotopy based algorithm to solve \eqref{eq:AM-step1} which can provide a solution with the desired cardinality $k_j$.
The homotopy algorithm for \eqref{eq:AM-step1} is itself an iterative scheme which starts with a zero vector and updates the solution and its support by adding or removing one element at a time \cite{OsbornePresnell_2000_NewApproachLasso,Efron_2004_LARS}. This way we can terminate the homotopy algorithm as soon as the cardinality of the homotopy solution becomes equal to $k_j$.\\
\textbf{Normalizing $x_j$: } The last step of Algorithm~\ref{alg:AM} normalizes the least squares solution $x_j$ to unit norm. Since the variables $x$ and $h$ are coupled together in the bilinear form \eqref{eq:y=Ax*h}, the magnitudes of the estimates $x_j$ and $h_j$ are inversely proportional to each other. At the same time, the optimization problem \eqref{eq:AM-step1} minimizes $\ell_1$ norm of $h_j$.  This reduces the magnitude of $h_j$ at every iteration and the magnitude of $x_j$ increases at the same rate. The normalization of either $x_j$ or $h_j$ brings a balance between their magnitudes. \\
\textbf{Stopping criteria: } We use the value of the residual and cardinality of channel estimate as an indicator for convergence. Some simple examples for the stopping criteria are: 1) Terminate Algorithm~\ref{alg:AM} after a fixed number of iterations, 2) Terminate when the cardinality $k_j$ of channel estimate $h_j$ reaches some threshold, or 3) Terminate when the residual $\|h_j \otimes Ax_j-y\|_2$ reduces below some threshold.\\
\textbf{Computational cost: } The main computational cost for Algorithm~\ref{alg:AM} involves solving \eqref{eq:AM-step1} and \eqref{eq:AM-step2} at every iteration. We can reduce the computational burden of solving these problems if we store QR factors for $A$ at the initialization and then use QR factors of $A$ and FFT operations to solve these problems. 

\section{Simulation}\label{sec:Results_ChannelPro}
In this section we present the simulation results to demonstrate the efficiency of the proposed random coding scheme.
The simulation setup is as follows. We start with a signal $x\in \mathbb{R}^n$ whose entries are drawn from standard normal distribution $N(0,1)$.
We use an $m\times n$ Gaussian matrix as the coding matrix $A$, whose entries are drawn from $N(0,1/m)$.
The received signal $y$ is generated according to the model \eqref{eq:y=Ax*h}, where $h \in \mathbb{R}^m$ is a $k$-sparse vector which is nonzero at randomly chosen $k$ locations. The nonzero entries of $h$ are also drawn from $N(0,1)$.
The signal $x$ and channel response $h$ is then estimated from $y$, by solving \eqref{eq:AM-x,h}, using alternating minimization scheme in Algorithm~\ref{alg:AM}.

\begin{figure}[t]
   \centering
   \subfigure[]{\includegraphics[width=1\columnwidth, scale = 1.0, angle=0]{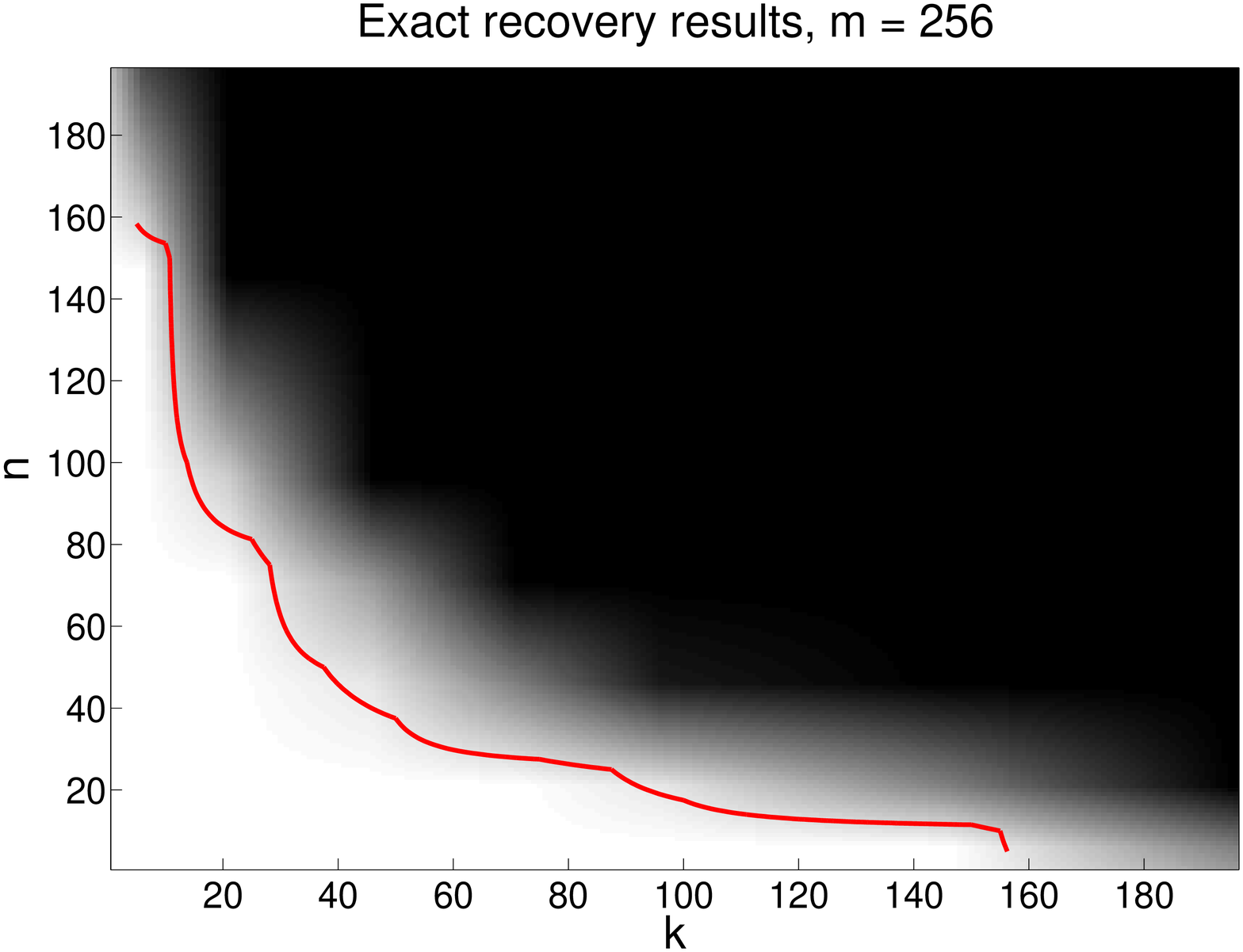}}
   \subfigure[]{\includegraphics[width=1\columnwidth, scale = 1.0, angle=0]{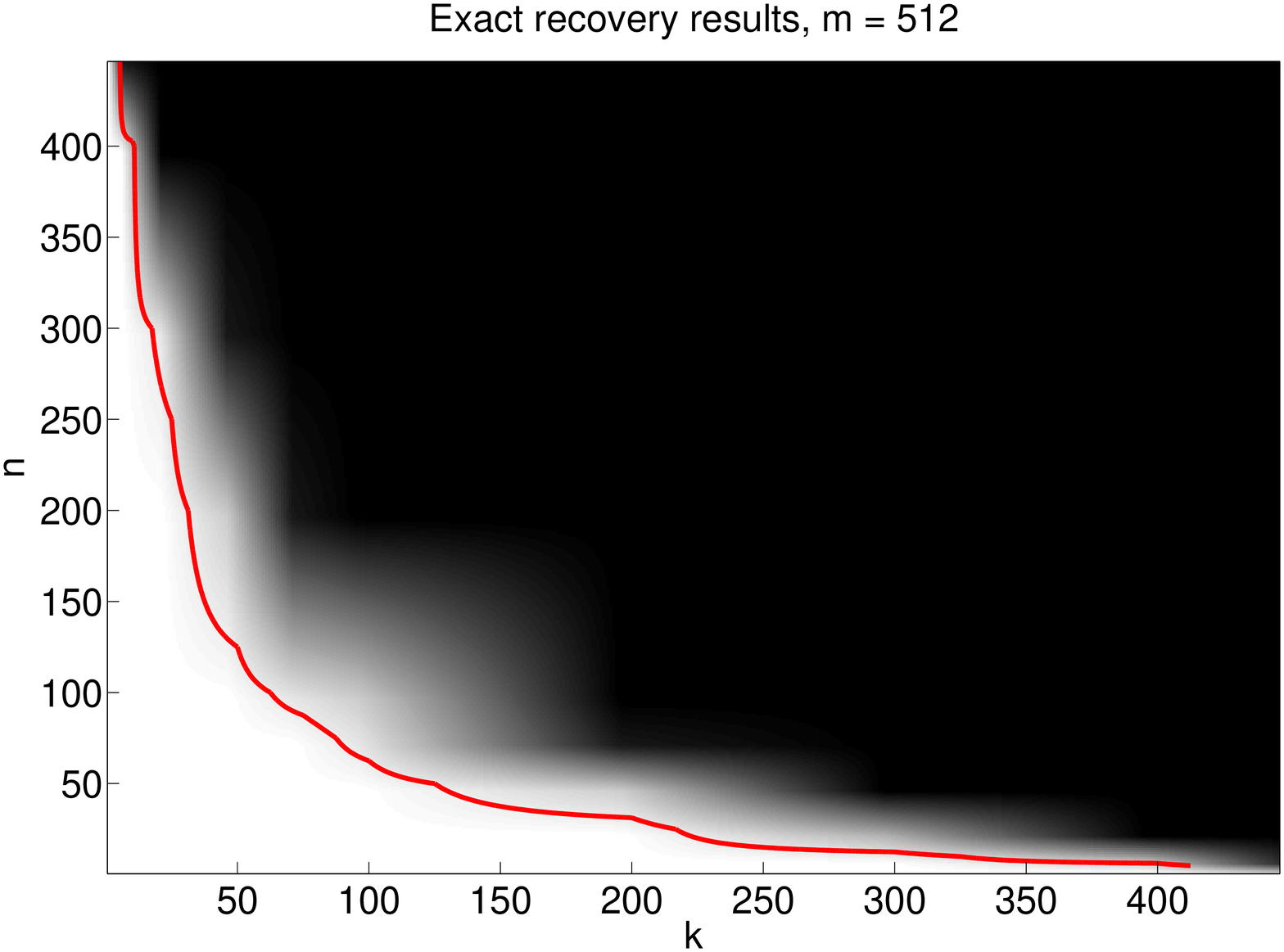}}
   \caption{Phase diagram for the performance of AM algorithm for (a) $m=256$ and (b) $m=512$. The intensity of the map gives the interpolated probability that the Algorithm~\ref{alg:AM} gives exact recovery for the respective values of $n$ and $k$, ranging from black (probability 0) to white (probability 1). The contour line at 95\% success rate is shown in solid (red).}
   \label{fig:Phase}
\end{figure}

We tested Algorithm~\ref{alg:AM} for the exact recovery of $x$ and $h$ with different values of $m$, $n$ and $k$. For every simulation, we determined if the error had converged to zero (with some small tolerance) within some preset number of iterations. 
In these simulations we assumed that we already know the location of the strongest component in the channel response $h$, instead of solving \eqref{eq:Initial_x0}.
The average recovery results for $m=256$ and $m=512$ are presented in the form of phase diagrams in Figure~\ref{fig:Phase}. We ran $10$ simulations on each $64$ locations of interest for the $m=256$ case and on $100$ locations for the $m=512$ case to produce these interpolated figures.

It can be seen from these results that if the underlying channel is sufficiently sparse, then we can identify the channel and estimate the signal correctly by using the random coding.

\section{Discussion}
We presented a random coding scheme to protect an arbitrary signal from the effects of multipath interference. We jointly estimate the channel and the transmitted signal by formulating the recovery problem as an optimization problem. We presented two recovery schemes in this regard -- block $\ell_1$ and alternating minimization.

In future, we want to develop some theoretical guarantees for when the recovery of $x$ and $h$ is possible in \eqref{eq:y=Ax*h}. As we discussed in section~\ref{sec:AM} that if we know either $x$ or $h$, we can solve \eqref{eq:AM-x,h} to recover the unknown variable exactly when $m \approx n+k$. However, these theoretical results do not apply when we do not know $x$ and $h$. We first want to determine the general conditions when the recovery of $x$ and $h$ is even possible. Then we want to explore the algorithms which can can reliably recover $x$ and $h$ under those conditions. AM algorithm is one promising option in this direction.

Note that our model in \eqref{eq:y=Ax*h} does not assume any noise in the received signal. A more realistic model for the received signal would be \[y = h \otimes Ax + \nu,\] where $\nu$ is a noise vector. The optimization problem \eqref{eq:AM-x,h} is robust to noise by its construction. Although we cannot recover $x$ and $h$ exactly from the noisy measurements here, but a good estimate can be made if noise level is not too large. The block $\ell_1$ method can be modified for the noisy measurements by replacing the equality constraint in \eqref{eq:Block-L1} with a {\em data fidelity} constraint, e.g., $\|\mathcal{A} \vec U - y \|_2 \le \gamma$ for some $\gamma > 0$ \cite{EldarMishali_2008_Robust}.

An important step in the AM algorithm is solving \eqref{eq:AM-step1}, at every iteration, for the desired cardinality of the solution. The homotopy scheme naturally lends itself to this iterative setup. Nonetheless, we can also use some other sparse recovery method, which can provide a solution with the desired cardinality. Two such methods are iterative thresholding \cite{Daubechies_2004_ItrativeThresholding,Blumensath-2008-IHT} and CoSaMP \cite{Needell_2008_CoSaMP}.


\bibliographystyle{ieeetr}
\bibliography{CS-bib}

\end{document}